\documentclass[preprint,showpacs,preprintnumbers,amsmath,amssymb]{revtex4}

\usepackage{graphicx}
\usepackage{dcolumn}
\usepackage{bm}

\begin{document}

\preprint{Phys. Rev. C 81, 052801(R)(2010)}

\title{
New Estimate for the Time-Dependent Thermal Nucleosynthesis of $^{180m}$Ta}

\author{T.~Hayakawa$^{1,2,*}$, T.~Kajino$^{2,3}$, S.~Chiba$^{2,4}$, G.~J.~Mathews$^{5}$}
\affiliation{
$^1$Kansai Photon Science Institute, Japan Atomic Energy Agency, Kizugawa, Kyoto 619-0215, Japan.\\
$^2$National Astronomical Observatory, Osawa, Mitaka, Tokyo 181-8588, Japan.\\
$^3$Department of Astronomy, Graduate School of Science, University of Tokyo, Tokyo 113-0033, Japan.\\
$^4$Advanced Science Research, Japan Atomic Energy Agency, Tokai, Naka, Ibaraki 319-11\\
$^5$Center for Astrophysics, Department of Physics, University of Notre Dame, Notre Dame, IN 46556}
\date{\today}

\begin{abstract}
We have made a new time-dependent calculation of  the supernova production ratio of the long-lived isomeric state  $^{180m}$Ta. Such a time-dependent solution is crucial for understanding the production and survival of this isotope.  We include the explicit linking between the isomer and all known excited states. We have also calculated the properties of possible links to a conjectured excited state which might decrease the final isomer residual ratio. We find that the  explicit time evolution of the synthesis of $^{180}$Ta using the available nuclear data avoids the overproduction relative to $^{138}$La for a $\nu$ process neutrino temperature of 4 MeV.

\end{abstract}

\maketitle

\begin{figure}
\includegraphics[viewport=0mm 0mm 200mm 170mm, clip, scale=0.8]{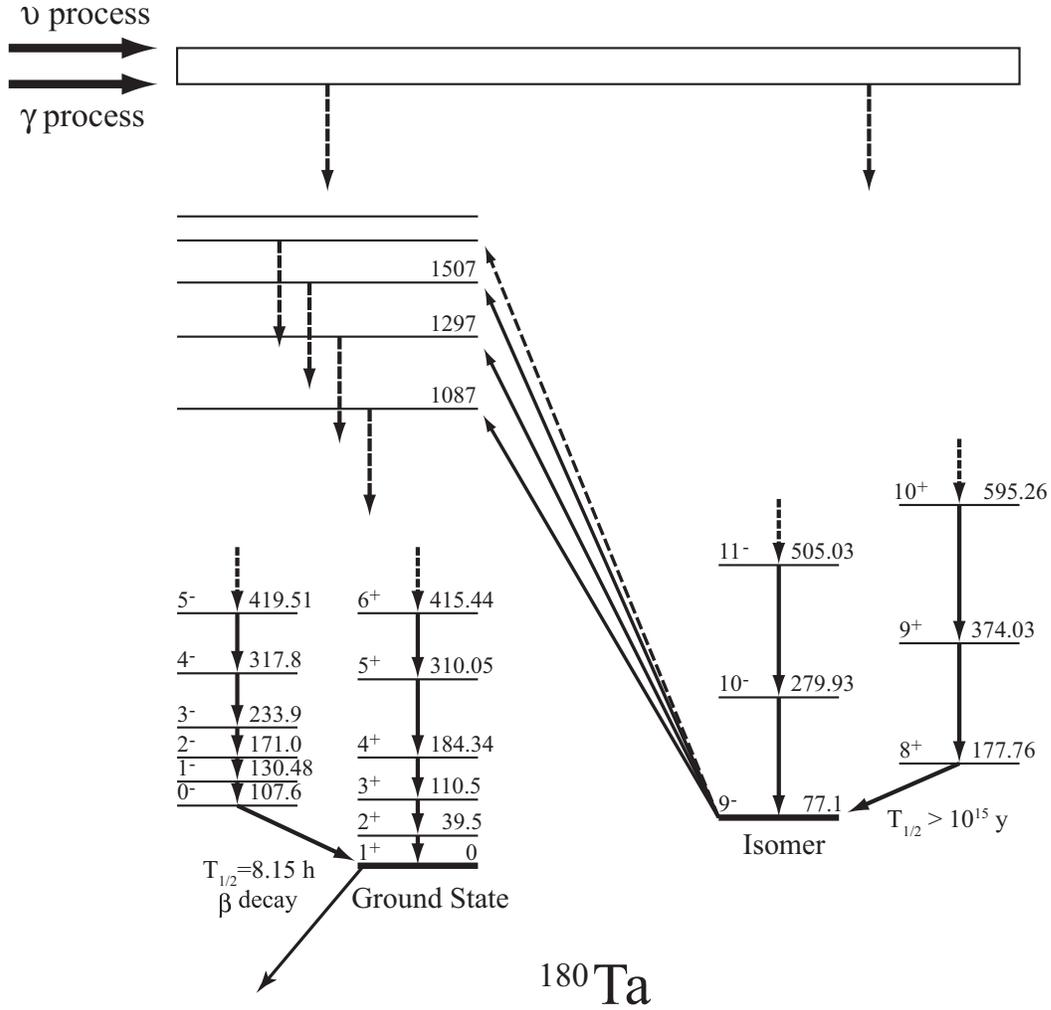}
\caption{
Partial nuclear level scheme of $^{180}$Ta. The ground state decays via $\beta$ decay with a half-life of 8.15 h,
while the isomer is a meta-stable state. The measured \cite{Belic02} excitation energies of states with weak transitions to the isomer are indicated.  These excited states should also connect to the ground state but their decay paths are unknown.
}
\label{fig:states}
\end{figure}

The nucleosynthesis of $^{180}$Ta has remained an unsolved problem.
For the most part this nucleus is bypassed by the major nucleosynthesis mechanisms of the $s$ and $r$ processes.  This accounts for why this isotope is the rarest in Nature.  
For this reason, a variety of more exotic processes,  
such as a weak branch through excited states in $^{180}$Hf \cite{Beer81},
the $\beta$ decay of $^{179}$Hf followed by neutron capture \cite{Yokoi83}, and the $\gamma$ process \cite{Woosley78,Arnould03,Hayakawa04,Hayakawa08}, have been proposed 
and studied experimentally \cite{Kellogg92,Belic99,Wisshak01}.
Perhaps, the most popular scenario in recent times is $^{180}$Ta production in  the $\nu$ process  \cite{Woosley90, Heger05} via the $^{181}$Ta$(\nu,{\nu}'n)^{180}$Ta and  $^{180}$Hf(${\nu}_e$,e)$^{180}$Ta neutrino reactions in core-collapse supernovae. 
It is currently believed  that only two isotopes ($^{138}$La and $^{180}$Ta) among the heavy elements
may be predominantly synthesized by the $\nu$ process \cite{Heger05}.
Although the new calculated result based upon results of a recent nuclear experiment \cite{Byelikov07} 
can reproduce the Solar abundance of $^{138}$La 
with charged current reactions and an electron neutrino temperature of 4 MeV, it overproduces
the abundance of $^{180}$Ta.
Here we investigate the possibility that
this overestimate originates from the unique feature that the naturally occurring abundance of $^{180}$Ta is actually a meta-stable isomer (half-life of $\ge$ 10$^{15}$ yr), 
while the true ground state is a $1^+$  unstable state which $\beta$-decays with a  half-life of only 8.15 hr (see Fig.~\ref{fig:states}).  
Therefore, a crucial ingredient for all of the possible production scenarios
is the ratio of the population of the meta-stable isomer to the total production of this isotope.

In the $\nu$ process, low-spin excited states in $^{180}$Ta 
are strongly populated from $^{180}$Hf by Gamow-Teller transitions and subsequently decay preferentially to the 1$^+$ ground state \cite{Heger05} (a similar situation for $^{138}$La is discussed in Ref.~\cite{Hayakawa08b}). However, in a high temperature photon bath,
the meta-stable isomer is excited from the ground state by ($\gamma$,$\gamma$') reactions through highly excited states.
Moreover, the transition rate between the ground state and the isomer 
is affected by the changing temperature.
Therefore, the final isomeric branching ratio should be evaluated by a time-dependent calculation.

Previous studies have noted \cite{Heger05, Byelikov07} that the observed $^{180m}$Ta abundance can not be inferred from their calculations until the branching between the long-lived isomer and the ground state is known.
If one knew sufficient  information on transitions between the two states of $^{180}$Ta,
the branching ratio could be  calculated by the  mathematical method  of \cite{Gupta01}. 
However, although many experiments have been carried out 
to identify the paths between
the two states  in $^{180}$Ta \cite{Dracoulis98,Saitoh99,Dracoulis00,Wheldon00,Wendel01},
such paths have never been identified and only transition probaiblities 
have been  measured \cite{Belic99,Belic02}.
In this letter,  therefore, we develop a  new calculation method  which can be applied even with limited information.
This method is then applied to provide a realistic estimate of the  branching in $^{180}$Ta. 
We show that this model leads to a good agreement between the final calculated 
isomeric abundance and the observed Solar-system abundance for a broad range 
of astrophysical parameters.

We adopt an exponential decrease $T$ = $T_{0}\exp{(-t/{\tau})}$ as a reasonable approximation to the adiabatic expansion of  shock heated material in supernovae \cite{Woosley78}.  Nuclei are completely thermalized in the high temperature regime so that the population ratio of  any  two states is simply given by a quotient of their Boltzman factors, i.e. 
$m_{i}/m_{j} = (2J_{i}+1)/(2J_{j}+1)\exp{[-(E_{i}-E_{j})/kT]}$,
where $m_{i}$ denotes the population of the state of $i$, with spin $J_i$,
and excitation energy, $E_i$.
For $T_9$ = 0.1$\--$1.0 (where $T_{9}$ is the temperature in units of 10$^9$ K)
all excited states lower than a few hundred keV 
are populated.
After the freeze-out each excited state decays to either the ground state or the isomer
and must be considered \cite{Mohr07}.

\begin{figure}
\includegraphics[viewport=0mm 0mm 200mm 120mm, clip, scale=1.0]{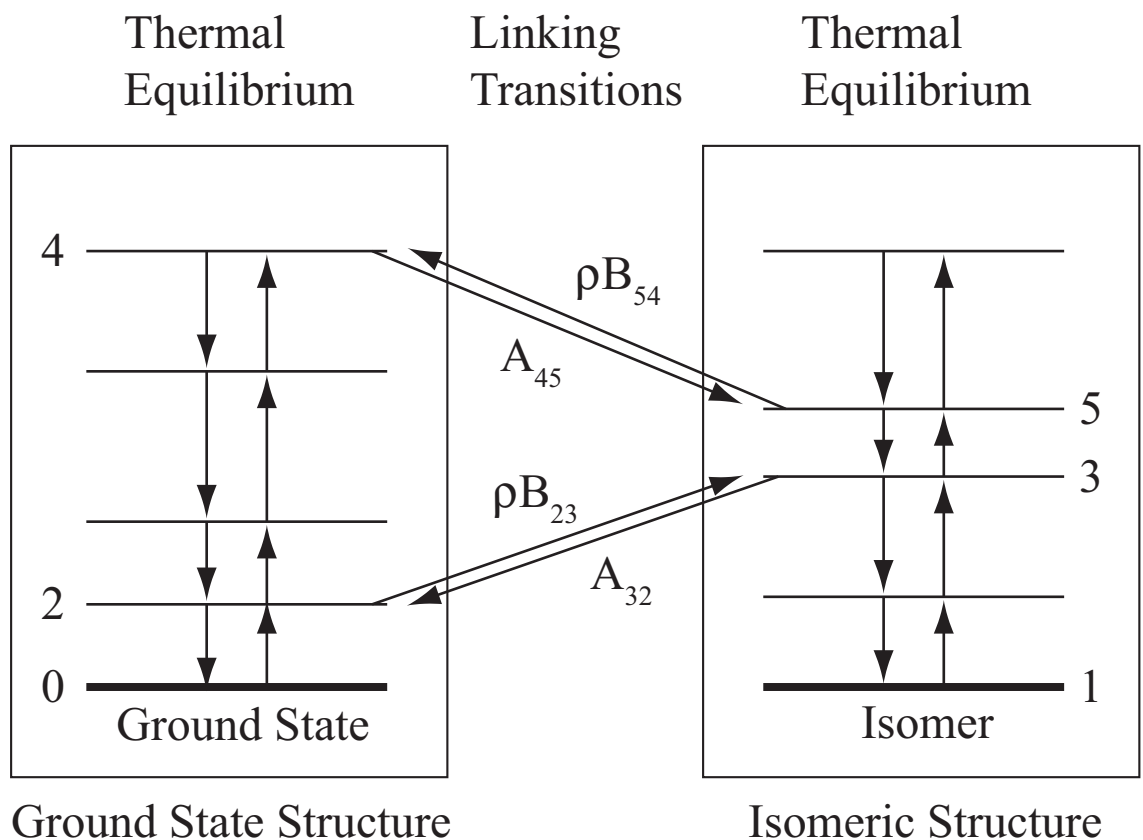}
\caption{
Schematic illustration of the nuclear structure relevant 
during the transitional temperature region.
The ground state structure (i.e. the ground state and excited states above the ground state)
is in thermal equilibrium.
The isomeric structure is also in thermal equilibrium.
The ground state and isomeric structures are connected via the indicated linking transitions.
}
\label{fig:ratio}
\end{figure}

In the transitional region, 
strongly connected states are only partly thermalized.
For our purposes we can model the excited-state structure of a deformed nucleus as consisting  of two sets of nuclear states:
1) the ground state structure, which consists of the ground state 
plus the excited states with strong transitions to the ground state;  and 2) the analogous isomeric structure.
The transition probability between states of the two structures
depends upon their quantum number $K$ 
which is the projected component of the total angular
momentum along the nuclear symmetry axis \cite{Walker01}.
Transitions with $\Delta$$K$ $>$ $\Delta$$I$
are forbidden, where $\Delta$$I$ is the transition multipolarity.
In the case of $^{180}$Ta 
the ground state and the isomer  have $K$ = 1 and 9, respectively.
Thus, the two structures can only  communicate by weak linking transitions
and each structure 
can be considered independently thermalized (see Fig,~\ref{fig:ratio}).
We can therefore treat these structures as two independent nuclear species.

To construct a model for the time evolution we consider two simple cases of linking transitions as shown in Fig.~\ref{fig:ratio}.
In the first case there  is a single linking transition between states 2 and 3.
In this case, the time-dependent evolution of the population probability of the ground-state structure, 
$N_{0}$ = ${\sum}m_{i}^{g}$/(${\sum}m_{i}^{g}$+${\sum}m_{j}^{m}$),
is given by
\begin{equation}
dN_{0}/dt = -P_{2}^{g}{\rho}B_{23}N_{0} +P_{3}^{m}A_{32}(1-N_{0})~~,
\label{eq:basic}
\end{equation}
where $A_{32}$ and $B_{23}$ are the Einstein coefficients between the indicated states, $\rho$ is the photon density
of a thermal Plank distribution, and $P_{i}^{g(m)}$ is the normalized population ratio of the excited state, 
$P_{i}^{g(m)}$ = $m_{i}^{g(m)}/{\sum}m_{j}^{g(m)}$.
In the case of $kT$ $<<$ ($E_{2}$-$E_{1}$), the Einstein coefficients 
are related by
${\rho}B_{12}$ = ($g_{2}$/$g_{1}$)$A_{21}$$\exp{[-(E_{2}-E_{1})/kT]}$,
where $g_{i} = (2J_i+1)$ is the spin statistical factor,
and we obtain,
\begin{equation}
dN_{0}/dt = -({g_{3}}/{g_{0}})\exp{[-(E_{3}-E_{0})/kT]}P_{0}^{g}A_{32}N_{0} 
+ ({g_{3}}/{g_{1}})\exp{[-(E_{3}-E_{1})/kT]}P_{1}^{m}A_{32}(1-N_{0})~~.
\end{equation}
A similar expression exists for the case of a single linking transition between excited states 4 and 5 on Fig.~\ref{fig:ratio}.  It is straightforward to extend this to the general case of multi-linking transitions.
The time-dependence of the population probability of the ground state structure is then given by
\[
\frac{dN_{0}}{dt}=-{\sum_{i,p}}P_{i}^{g}A_{ip}N_{0}
+{\sum_{i,p}}P_{i}^{m}{\rho}B_{pi}(1-N_{0}),
-{\sum_{j,q}}P_{j}^{g}{\rho}B_{qj}N_{0}
+{\sum_{j,q}}P_{j}^{m}A_{jq}(1-N_{0})
\]
\begin{equation}
 = -{\sum_{i,p}}P_{0}^{g}\frac{g_{i}}{g_{0}}\exp{[-(E_{i}-E_{0})/kT]}A_{ip}N_{0}
+{\sum_{j,q}}P_{1}^{m}\frac{g_{j}}{g_{1}}\exp{[-(E_{j}-E_{1})/kT]}A_{jq}(1-N_{0}),
\label{eq:general}
\end{equation}
where, $0$ and $1$ denote the ground state and the isomer, respectively,
$i$ ($j$) denotes  levels of the ground state structure (or the isomer structure)
and $p$ ($q$) denotes the levels of any other structure.

In general, excited states of deformed nuclei are characterized by collective rotational motion. 
Each excited state is a member of a rotational band, in which the
electric transition
probabilities are enhanced by 1$\--$2 orders of magnitude.  However the interband transition probabilities
between two such rotational bands
are much hindered relative to  the intraband transitions. 
Therefore, in the case of deformed nuclei, the ${\Gamma}_{i}$ 
corresponding to the interband transition rates are much smaller than ${\Gamma}_{0}$
and one can approximate  $g_{i}$/$g_{1}$${\Gamma}_{i}$${\Gamma}_0$/${\Gamma} \approx g_{i}$/$g_{1}$${\Gamma}_{i}$.   Finally, inserting  $A$ = $\Gamma$/$\hbar$, into  equation (\ref{eq:general}), 
we obtain
\begin{equation}
\frac{dN_{0}}{dt}= -{\sum_{i}}P_{0}^{g}\frac{g_{1}}{g_{0}}\exp{[-(E_{i}-E_{0})/kT]}\frac{g_{i}}{g_{1}}\frac{{\Gamma}_{i}}{\hbar}N_{0}
+{\sum_{j}}P_{1}^{m}\exp{[-(E_{i}-E_{1})/kT]}\frac{g_{j}}{g_{1}}\frac{{\Gamma}_{j}}{\hbar}(1-N_{0}).
\label{eq:final}
\end{equation}

We have applied this general formula [Eq.~(\ref{eq:final})] to the case of $^{180}$Ta. 
The excitation energies and spins have been  taken from an evaluated data set of Ref.~\cite{Data03}.
We have calculated $P^{g(m)}_{i}$ 
by taking into account all known excited states up to  600 keV excitation energy.
The isomer population ratio $P_{1}^{m}$ = $m_{1}^{m}$/${\sum}m^{m}_{i}$
 is 0.85$\--$0.94 during the transitional temperature region 
of $T_9$ = 0.44$\--$0.62 (see below).
This result indicates that the linking transitions connected directly 
with the isomer are crucial for the total transition rate between the ground state and isomeric structures.
Note that the transition between states 2 and 3 in Fig.~\ref{fig:ratio} are negligibly small
since the spins of the excited states above the isomer are generally larger than the spin of the isomer ($J$ = 9).

\begin{figure}
\includegraphics[viewport=0mm 0mm 200mm 160mm, clip, scale=0.8]{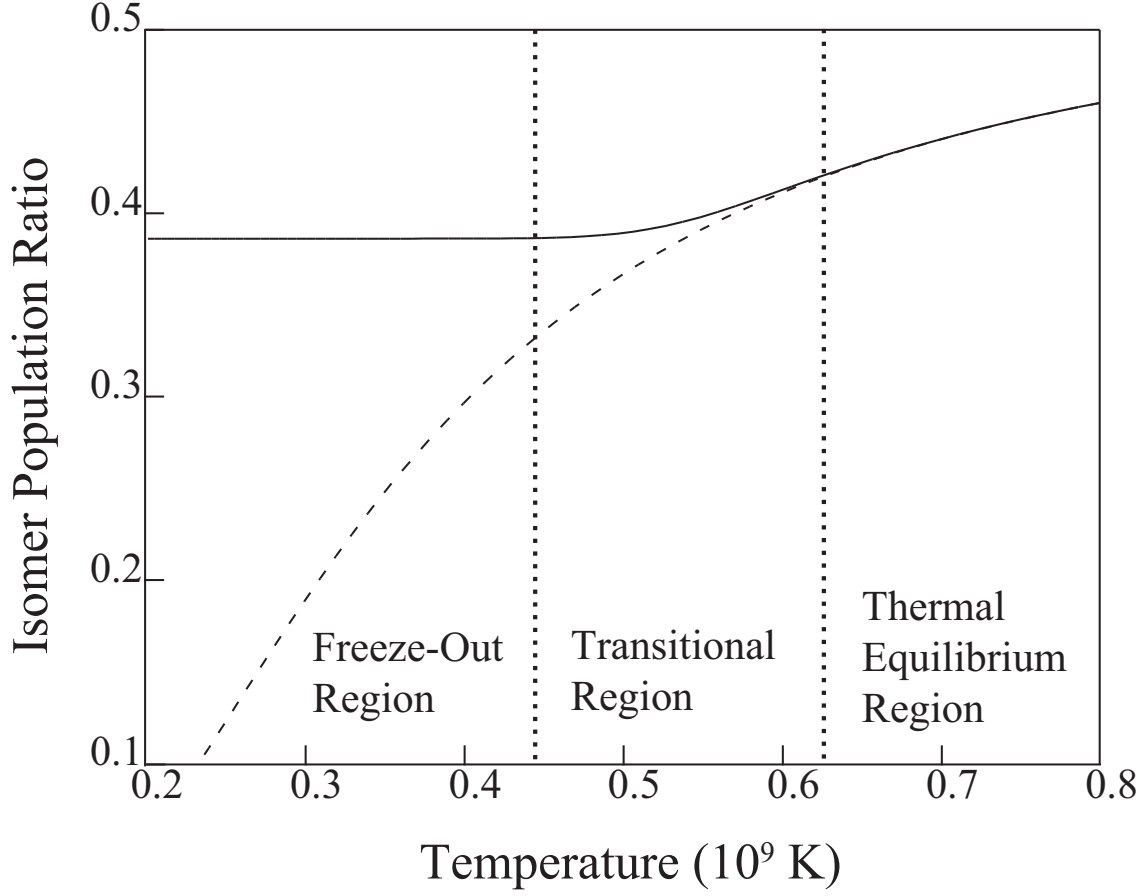}
\caption{
Calculated isomer population ratio. The solid line denotes the time-dependent calculated result.
The dashed line denotes the isomer ratio under  the condition of thermal equilibrium. In the high temperature regime ($T_9$ $>$ 0.62) both are identical. In the low temperature regime ($T_9$ $<$ 0.62) the calculated ratio is much higher than the thermal equilibrium ratio.
}
\label{fig:timeratio}
\end{figure}

Belic {\it et al.}~reported on the partial widths $g_{i}$/$g_{1}$${\Gamma}_{i}$${\Gamma}_0$/${\Gamma}$ [meV] \cite{Belic02} for 
9 transitions connected with the isomer.  However, 
 transitions between the ground state  and the excited states connected to the isomer 
 have not been identified (see Fig.~\ref{fig:states}).
Figure \ref{fig:timeratio} shows the calculated isomer population ratio taking into account the 9 transitions in $^{180}$Ta. 
For initial conditions we begin with static thermal equilibrium among states at $T_9$ = 1.0, and we take $\tau$ = 1 s for the supernova temperature time constant.
The population of the isomer structure decreases with decreasing temperature as expected. 
In the high temperature region ($T_9 >$ 0.62) the present calculated result is identical 
with that obtained by assuming  static thermal equilibrium. 
However, in the low temperature region ($T_9 <  0.62$) the present time-dependent calculated result is significantly different from
the thermal equilibrium result.
In the freeze-out region at low temperature ($T_9 <  0.44$)
the two structures are completely disconnected and the isomer population ratio remains fixed at 
$P_{m}$/($P_{m}$+$P_{gs}$) = 0.39 $\pm$ 0.01. The uncertainty which is evaluated from the experimental errors of the energy width ${\Gamma}_{i}$ \cite{Belic02} is small since both the transition rates of m $\rightarrow$ gs and gs $\rightarrow$ m are proportional to ${\Gamma}_{i}$.

\begin{figure}
\includegraphics[viewport=0mm 0mm 210mm 170mm, clip, scale=0.7]{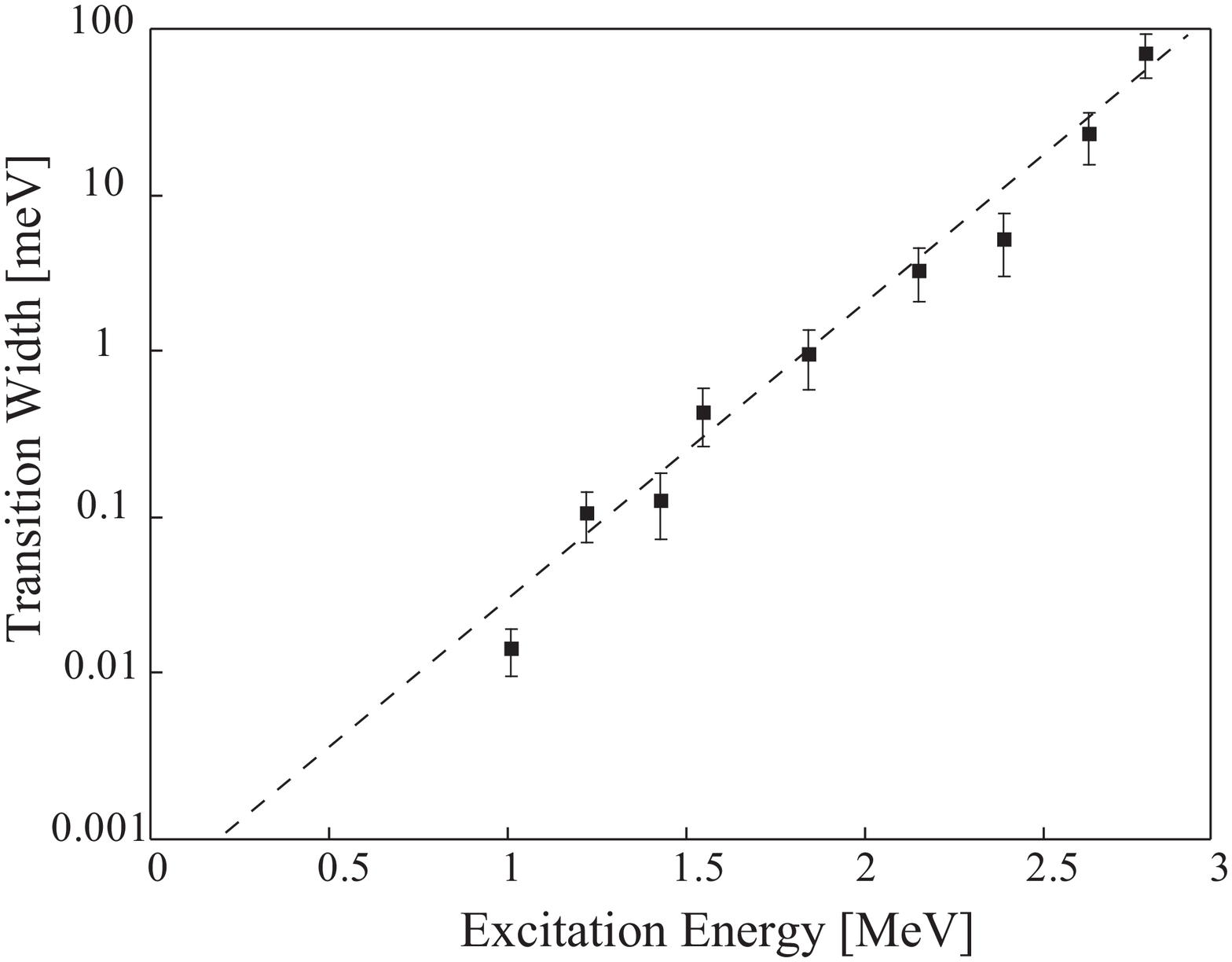}
\caption{
Observed transition widths $g_{i}$/$g_{1}$${\Gamma}_{i}$${\Gamma}_0$/${\Gamma}$ [meV] of the 9 known linking transitions as a function of their excitation energy. 
The Data are taken from Ref.~\cite{Belic02}.
The dashed line is a simple exponential least square fit to these widths used to estimate widths for other hypothetical linking transitions.
}
\label{fig:obsratio}
\end{figure}

For completeness we should also remark that it has been suggested \cite{Mohr07} 
that an unobserved linking transition to a state at  592 keV  
might also exist.
To estimate the transition probability, we plot in Fig.~\ref{fig:obsratio} the 9 known linking transition widths
to the isomer as a function of their excitation energy. 
The transition widths increase with  excitation energy and can be fit with a simple exponential growth.
This trend  can be understood in terms of a $K$ mixing effect caused by the Coriolis interaction,
but a detailed explanation of why a simple exponential growth reproduces the observed transition probability so well is beyond  the scope of this Letter.  Nevertheless, from a least-square fit to this trend (dashed line on Fig.~\ref{fig:obsratio}) we can estimate that an unknown state at 592 (800) keV would have a width of 
 $g_{i}$/$g_{1}$${\Gamma}_{i}$ = 0.003 (0.008) [meV].

The isomeric residual population ratio is then $P_{m}$/($P_{g}$+$P_{m}$) = 0.18 and 0.30 
for hypothetical additional linking transitions to states at 592 keV and 800 keV, respectively.
The isomer ratio is sensitive to the lowest energy of the linkning transitions.
Note, however, that available experimental results for this nucleus have found no evidence for direct transitions below 1.09 MeV \cite{Belic02} or 0.739 MeV  \cite{Lakosi02},
and that previous $\gamma$-ray spectroscopy experiments \cite{Dracoulis98,Saitoh99,Dracoulis00,Wheldon00,Wendel01} 
have measured no semi-direct transitions lower than 1.09 MeV;
this fact indicates that the lowest energy of the linking transitions is the known energy
of 1.09 MeV.
We choose, therefore, $P_{m}$/($P_{g}$+$P_{m}$) = 0.39
based only on the  observed 9 linking transitions for the following discussion.

In the $\nu$ and $\gamma$ processes, $^{180}$Ta is initially synthesized in environments with $T_9$ $>$ 0.62 so that $^{180}$Ta is completely thermalized 
and the production ratio in any previous nuclear reaction process 
does not affect the final relative isomeric population. 
The isomer population at freeze-out is insensitive to the temperature
 time constant because  of the slow rate 
of change of the population near  the freeze-out.
The calculated values are almost identical for temperature timescales in the range of $\tau$ = 0.3$\--$3 s.
As we shall now show, the only remaining sensitive astrophysical parameter is the neutrino temperature for the $\nu$ process.

Byelikov et al. \cite{Byelikov07} presented nucleosynthesis yields 
and it was concluded that $^{138}$La was reproduced by neutrino charged current reactions and 
$\nu_{e}$ temperatures around 4 MeV.
However, $^{180}$Ta was overproduced without correcting of the isomer residual ratio.
In Table 1, we present the modified yields taking into account the isomer residual ratio.
The quoted production ratios of $^{138}$La and $^{180}$Ta are normalized to $^{16}$O which is known to be predominantly produced in supernovae and its solar abundance is very large compared with that of the heavy elements.  A successful nucleosynthesis mechanism must  produce abundances relative to $^{16}$O near unity.  Otherwise, these isotopes are under (or over) produced.  Both $^{138}$La and $^{180}$Ta are now produced in relative amounts near unity 
by the charged current neutrino reactions and a ${\nu}_{e}$ temperature of 4 MeV.
Although this result is only from the 15 M$_\odot$ model the core properties do not depend much upon progenitor mass and the initial mass function is weighted toward low-mass progenitors.  Hence the  15 M$_\odot$ model should be a good representation of a proper average over supernova progenitors.  

We stress that, in the present calculation,  the isomer ratio is almost 
independent of the astrophysical parameters such as
the peak temperature, the temperature time constant, 
the supernova neutrino energy spectrum, and the explosion energy.
The final result that the overproduction 
problem of $^{180}$Ta in the $\nu$ process is largely reduced 
is probably robust,
though only a full set of stellar nucleosynthesis 
calculations that include the time-dependent depopulation of the 
isomer will answer this question quantitatively.
As shown in Table \ref{table:yields}, 
the present result constrains the neutrino energy spectrum, which is of importance
 for understanding the supernova explosion mechanim and the detection of  supernova neutrinos in near future. 
A previous study \cite{Yoshida06} has shown that light element yields of the $\nu$ process depend upon the  neutrino oscillation parameter ${\Theta}_{13}$. 
Heger {\it et al.} suggested that the yields of $^{138}$La and $^{180}$Ta is sensitive to
${\Theta}_{13}$ \cite{Heger05}.
Using the present isomer ratio, 
one can now study systematically the neutrino oscillation effect for both of light and heavy elements.

This work has been supported in part by Grants-in-Aid for Scientific
Research (21340068, 20244035, 20105004) of Japan.   
Work at the University of Notre Dame (G.J.M.) supported
by the U.S. Department of Energy under 
Nuclear Theory Grant DE-FG02-95-ER40934.

\begin{table}
\caption{
Nucleosynthesis production factor relative to Solar (normalized to $^{16}$O)
for various $\gamma$ or $\nu$ process conditions  (column 1) based upon the 15 M$_\odot$ supernova model of Ref.~\cite{Heger05}.  In the notation of Ref.~\cite{Byelikov07}, + n.c.~means the $\gamma$ process plus neutral-current interactions, and ++c.c.~denotes that charged current interactions are also included at the indicated neutrino temperatures.  Column 2 gives the total $^{138}$La  yields and column 4 gives the total (g.s.~+~isomer) $^{180}$Ta yields \cite{Heger05,Byelikov07}.  The yields of $^{180m}$Ta (column 3) deduced in the present work  are based upon  our calculated isomer residual population fraction of 0.39. Note that for $T_\nu = 4$ MeV $^{138}$La and $^{180m}$Ta are produced with about the same over-production factor.}
\begin{tabular}{lccc} 
\hline
   Model              & $^{138}$La  & {\bf $^{180m}$Ta} &  $^{180}$Ta(g.s.+isomer) \\
\hline
$\gamma$-proc. only  &  0.190   &   {\bf  0.234}  & 0.599 \\
+ n.c. (6 MeV)       &  0.280   &   {\bf  0.399}  & 1.024 \\
++ c.c. (4 MeV)      &  1.101   &   {\bf  1.218}  & 3.123  \\
++ c.c. (6 MeV)      &  2.797   &   {\bf  2.109}  & 5.409 \\
++ c.c. (8 MeV)      &  3.222   &   {\bf  2.881}  & 7.387 \\
\hline
\end{tabular}
\label{table:yields}
\end{table}

$^{*}$ Electronic address: hayakawa.takehito@jaea.go.jp

\newpage

\begin{thebibliography}{99}


\bibitem{Beer81} H. Beer and R. A. Ward, Nature {\bf 291}, 308 (1981).
\bibitem{Yokoi83} K. Yokoi and K. Takahashi, Nature {\bf 305}, 198-201 (1983).
\bibitem{Woosley78} S.E. Woosley,  W.M. Howard, Astrophys. J. Suppl. {\bf 36}, 285 (1978).
\bibitem{Arnould03}  M. Arnould, and S. Goriely, Phys. Rep. {\bf 384}, 1 (2003).
\bibitem{Hayakawa04} T. Hayakawa,  N. Iwamoto, T. Shizuma, T. Kajino, H. Umeda, K. Nomoto, Phys. Rev. Lett. {\bf 93}, 161102 (2004).
\bibitem{Hayakawa08} T. Hayakawa, N. Iwamoto, T. Kajino, T. Shizuma, H. Umeda, K. Nomoto, Astrophys. J. {\bf 685}, 1089 (2008).
\bibitem{Kellogg92} S.E. Kellogg, and E.B. Norman, Phys. Rev. C {\bf 46}, 1115 (1992).
\bibitem{Belic99} D. Belic, et al., Phys. Rev. Lett. {\bf 83}, 5242 (1999).
\bibitem{Wisshak01} K. Wisshak et al., Phys. Rev. Lett. {\bf 87}, 251102 (2001).

\bibitem{Woosley90} S.E. Woosley, D.H. Hartmann, R.D. Hoffman and W.C. Haxton, Astrophys. J. {\bf 356}, 272 (1990).
\bibitem{Heger05} A. Heger E. Kolbe, W.C. Haxton, K. Langanke, G. Mart{\'i}nez-Pinedo, S.E. Woosley, Phys. Lett. {\bf B606}, 258 (2005).
\bibitem{Byelikov07} A. Byelikov, et al., Phys. Rev. Lett. {\bf 98}, 082501 (2007).
\bibitem{Hayakawa08b} T. Hayakawa, T. Shizuma, T. Kajino, K. Ogawa, H. Nakada, Phys. Rev. C {\bf 77}065802 (2008).

\bibitem{Gupta01} S.S. Gupta and B.S. Meyer, Phys. Rev. C {\bf 64}, 025805 (2001).

\bibitem{Dracoulis98} G.D. Dracoulis, {\it et al}.,  Phys. Rev. C {\bf 58}, 1444 (1998).
\bibitem{Saitoh99} T.R. Saitoh, {\it et al.} Nucl. Phys. {\bf A660}, 121 (1999).
\bibitem{Dracoulis00} G.D. Dracoulis, T.Kibe{\`d}i, A.P. Byrne, R.A. Bark, and A.M. Baxter, Phys. Rev. C {\bf 62}, 037301 (2000).
\bibitem{Wheldon00} C. Wheldon {\it et al.}, Phys. Rev. C {\bf 62}, 057301 (2000).
\bibitem{Wendel01} T. Wendel, {\it et al.}, Phys. Rev. C {\bf 65}, 014309 (2001).

\bibitem{Belic02} D. Belic, {\it et al.,} Phys. Rev. C {\bf 65}, 035801 (2002).
\bibitem{Walker99} P. Walker, and G. Dracoulis, Nature, {\bf 399}, 25 (1999).



\bibitem{Mohr07} P. Mohr, F. K{\"a}ppeler, R. Gallino, Phys. Rev. C {\bf 75}, 012802(R) (2007).
\bibitem{Walker01} P. M. Walker, G. D. Dracoulis, J. J. Carroll, Phys. Rev. C {\bf 64}, 
061302(R) (2001).
\bibitem{Data03} S.-C. Wu, and H. Niu, Nucl. Data Sheets, {\bf 100}, 483 (2003).
\bibitem{Lakosi02} L. Lakosi \& T. C. Nguyen, Nucl. Phys. {\bf A697}, 44 (2002).
\bibitem{Yoshida06} T. Yoshida {\it et al.,} Phys. Rev. Lett. {\bf 96}, 091101  (2006).

\end{thebibliography}
\end{document}